\documentclass[pra,aps,twocolumn,showpacs,showkeys]{revtex4}
\usepackage{amsmath}

\begin{document}

\title{Entanglement Purification of Any Stabilizer State}
\author{S. Glancy}
\email{sglancy@boulder.nist.gov}
\author{E. Knill}
\email{knill@boulder.nist.gov}
\affiliation{Mathematical and Computing Science Division, Information Technology Laboratory, National Institute of Standards and Technology, Boulder, Colorado 80301, USA}
\author{H. M. Vasconcelos}
\email{Hilma.M.DeVasconcelos.2@nd.edu}
\affiliation{Department of Physics, University of Notre Dame, Notre Dame, Indiana 46556, USA}
\date{\today}
\begin{abstract}
We present a method for multipartite entanglement purification of any
stabilizer state shared by several
parties. In our protocol each party measures the stabilizer operators
of a quantum error-correcting code on his or her qubits.  The parties
exchange their measurement results, detect or correct errors, and
decode the desired purified state.  We give sufficient
conditions on the stabilizer codes that may be used in this procedure
and find that Steane's seven-qubit code is the smallest
error-correcting code sufficient to purify any stabilizer state. An
error-detecting code that encodes two qubits in six can also be used
to purify any stabilizer state. We further specify which classes of
stabilizer codes can purify which classes of stabilizer states.
\end{abstract}
\maketitle

\section{Introduction}

In this paper we describe a method for entanglement purification that
is able to purify any stabilizer or graph state.  The goal of a
purification protocol is to increase the purity of a mixed state whose
qubits are divided among several parties able to communicate only
through classical channels.  Specifically, suppose the state
$|\psi\rangle$ is a pure entangled state of $n$ qubits and each qubit
of $|\psi\rangle$ is sent to a different party (Alice, Bob, Charlie,
\ldots). In this initial stage, the qubits are transmitted through
noisy quantum channels, so there is some probability that each may be
affected by Pauli $\sigma_x$, $\sigma_y$, or $\sigma_z$
errors. After the transmission, $|\psi\rangle$ becomes the mixed state
$\hat{\rho}$.  If $m$ copies of $|\psi\rangle$ are prepared and
transmitted, so that each party holds $m$ qubits, it may be possible
to obtain one or more copies of $|\psi\rangle$ with less noise (more
purity). For this purpose, the parties may use local quantum
operations and measurements on their own qubits and classical
communication.

The entanglement purification (or ``entanglement distillation'')
problem was first studied by Bennett and co-authors in 1995
\cite{Bennett95}.  They describe a method for two separate parties to
purify a Bell pair by use of local operations on copies of noisy Bell
pairs and classical communication between the two parties.  Since then
many other researchers have studied this problem, introducing new
protocols for the purification of Bell pairs and providing methods for
purifying other classes of entangled states. For examples of this
research see
\cite{Bennett96,Maneva00,Matsumoto02,Aschauer04,Hostens04,Watanabe05,Miyake05,Hostens05,Cheong05,Kruszynska06}.

In the remainder of this section we give a brief review of stabilizers
and introduce our method of using operator arrays to characterize many
copies of entangled states shared by sevaral parties. In
Sec.~\ref{errorcorrectingcodesforpurification} we discuss the use of
quantum error-detecting and -correcting codes for purification. In
Sec.~\ref{statesandcodes} we give classes of stabilizer codes that can
be used to purify specific classes of states including the class of
all stabilizer states. In Sec.~\ref{conclusion} we discuss these
results and make some concluding remarks.

\subsection{Stabilizer Review}

When using the stabilizer formalism, instead of writing out vectors in
a Hilbert space, we specify a quantum state $|\psi\rangle$ using (i) a
set of operators that have $|\psi\rangle$ as an eigenstate and (ii)
the eigenvalues of $|\psi\rangle$ under each such operator (see
chapter 10 of \cite{Nielsen00}). This set of operators is the
stabilizer of $|\psi\rangle$, and we say ``$|\psi\rangle$ is
stabilized by'' this set. Although the stabilizer often refers to a
set of operators whose eigenvalue is one, in this paper we allow the
stabilizer to include operators with other eigenvalues. Several states
may share the same stabilizer, but have different eigenvalues for the
members of the stabilizer.  We are interested in states determined by
stabilizers consisting of the set of stabilizing Pauli products --
tensor products of Pauli matrices including the identity matrix. Such
states are called ``stabilizer states''.  Stabilizer states are
equivalent to ``graph states''~\cite{Schlingemann01}, where the
latter are specified by means of graphs rather than stabilizers.  The
stabilizer of stabilizer states necessarily consists of commuting
Pauli products. It forms a projective group (closed under
multiplication up to a phase), so it is sufficient to specify its
generators.  The number of independent generators must equal the
number of qubits. Because Pauli matrices have eigenvalues $\pm 1$,
each generator must also have eigenvalue $\pm 1$.

For example, the Bell state $|B_{00}\rangle=|00\rangle+|11\rangle$
(with normalization omitted) has stabilizer generators $XX$ and $ZZ$
and eigenvalues $+1$ and $+1$, respectively. Here we use the
abbreviations $X$, $Y$ and $Z$ for the Pauli $\sigma_x$, $\sigma_y$
and $\sigma_z$ matrices. Expressions such as $XY$ refer to the Pauli
product where $\sigma_x$ and $\sigma_y$ act on the first and second
qubits, respectively.  We can specify $|B_{00}\rangle$ by means of 
its stabilizer and eigenvalues as follows:
\begin{equation}
|B_{00}\rangle=\left[\begin{array}{cc}
X &X\\
Z &Z
\end{array}\right]
,
\left[
\begin{array}{c}
+1 \\
+1
\end{array}
\right].
\end{equation}
Note that these are not numeric matrices in the square brackets.  We
are simply listing each generator and its corresponding
eigenvalue on each row.  The stabilizers of the three other Bell
states are also generated by $XX$ and $ZZ$, but they have different
eigenvalues.  The three qubit Greenberger-Horne-Zeilinger (GHZ)
\cite{Greenberger} state is
\begin{equation}
|000\rangle+|111\rangle=\left[\begin{array}{ccc}
X&X&X\\
Z&Z&I\\
Z&I&Z
\end{array}\right]
,
\left[
\begin{array}{c}
+1 \\
+1 \\
+1
\end{array}
\right].
\end{equation}
The Bell states and the GHZ states are both in the class of CSS states
(named after Calderbank, Shore and Steane). A stabilizer state is a
CSS state if it can be transformed with unitary single qubit
operations into a form in which each of its stabilizer generators can
be written using only $X$'s and $I$'s or $Z$'s and $I$'s (the ``CSS
form''). The class of CSS states is equivalent to the two-colorable
graph states \cite{Chen04}. Methods for purifying any CSS state are
already known \cite{Aschauer04, Hostens05}.

Consider the state
\begin{equation}
|\Delta\rangle=\left[\begin{array}{ccc}
X & Z & Z \\
Z & X & Z \\
Z & Z & X
\end{array}\right],
\left[\begin{array}{c}
+1 \\
+1 \\
+1
\end{array}\right].
\end{equation}
This state cannot be written in the CSS form, and it is impossible to
transform $|\Delta\rangle$ into a CSS state by using single qubit
operations. Readers familiar with the techniques of graph states may
recognize that this state has a triangle graph.

We can also use the stabilizer formalism to describe quantum
error-detecting and -correcting stabilizer codes.  (Since all codes
considered here are stabilizer codes, from now on we omit the
modifier ``stabilizer''.)  To define a code, the stabilizer generators
are used to specify a (more than one dimensional) subspace into which
one may encode quantum information. In this case, the subspace
consists of the states with identical eigenvalues for each generator.
If we want to encode $m$ logical qubits using $n$ physical qubits, the
code is specified by $n-m$ independent generators of the stabilizer.
For example the four qubit error-detecting code $\mathcal{C}_4$
encodes two logical qubits using four physical qubits
\cite{Gottesman97} and is described by the stabilizer with generators
\begin{equation}
\mathcal{C}_4 \Leftrightarrow \left[\begin{array}{cccc}
X & X & X & X \\
Z & Z & Z & Z 
\end{array}\right],
\left[\begin{array}{c}
+1 \\
+1
\end{array}\right].
\end{equation}
The Hilbert space of the four physical qubits contains 16
dimensions, but the set of states with $+1$ eigenvalues of the above
operators is a four-dimensional subspace.  Four dimensions are
sufficient to contain two logical qubits, which we specify by giving
their logical Pauli operators:
\begin{subequations}
\begin{eqnarray}
X_L^{(1)} & = & XXII \label{c4logical1}\\
Z_L^{(1)} & = & ZIZI \label{c4logical2}\\
X_L^{(2)} & = & IXIX \label{c4logical3}\\
Z_L^{(2)} & = & IIZZ.\label{c4logical4}
\end{eqnarray}
\label{c4logical}
\end{subequations}
The encoded logical operators have all the algebraic relationships
that we expect from the $X$ and $Z$ operators on two qubits. Because
the state of this four qubit system is always confined to the $+1$
eigenstate of the stabilizers, there are many equivalent
representations of the logical qubit operators, which we can obtain by
multiplying the logical operators by members of the
stabilizer group. For example we can also write
\begin{equation}
X_L^{(1)}\simeq XXXX \times XXII\simeq IIXX,
\end{equation}
because each of these alternatives has the same effect on the
encoded logical qubits. 

$\mathcal{C}_4$ is an error-detecting code. An $X$ error on
any single physical qubit results in a state for which the eigenvalue
of the generator $ZZZZ$ is changed to $-1$. A $Z$ error will change
the eigenvalue of $XXXX$ to $-1$, and a $Y$ error changes both
eigenvalues.  To detect errors we simply measure each of the code's
generators. We call each such measurement a ``parity check'', and the
operator being measured is the ``parity check operator''. We assume
that the measurements are projective so that the state after
a measurement is a projection of the initial state onto one of the two
eigenspaces of the parity check operator.  From each parity check we
obtain an eigenvalue, which must be $\pm 1$.  The ``syndrome'' is the
vector of eigenvalues for the generators.  The eigenvalue for any
parity check operator can be inferred from the syndrome.  Because for
$\mathcal{C}_4$, the syndrome does not tell us which qubit received
the error, we are unable to correct one-qubit errors.

Codes such as $\mathcal{C}_4$ with the property that there is a choice
of stabilizer generators that can be written with only $X$'s and
$I$'s or $Z$'s and $I$'s are called ``CSS codes''.

The smallest error-correcting code that protects a single logical
qubit requires five physical qubits \cite{Bennett96,Laflamme96}. It
has the generators
\begin{equation}
\mathcal{C}_5 \Leftrightarrow \left[\begin{array}{ccccc}
X & Z & Z & X & I \\
I & X & Z & Z & X \\
X & I & X & Z & Z \\
Z & X & I & X & Z
\end{array}\right],
\left[\begin{array}{c}
+1 \\
+1 \\
+1 \\
+1 
\end{array}\right].
\end{equation}
This code has the logical operators
\begin{subequations}
\begin{eqnarray}
X_L & = & XXXXX \\
Z_L & = & ZZZZZ,
\end{eqnarray}
\end{subequations}
and is not a CSS code.

Another code that we use is Steane's seven qubit code
$\mathcal{C}_7$, which encodes one logical qubit in seven physical
qubits \cite{Steane96}. It has the generators
\begin{equation}
\mathcal{C}_7 \Leftrightarrow \left[\begin{array}{ccccccc}
X & I & X & I & X & I & X \\
I & X & X & I & I & X & X \\
I & I & I & X & X & X & X \\
Z & I & Z & I & Z & I & Z \\
I & Z & Z & I & I & Z & Z \\
I & I & I & Z & Z & Z & Z \\
\end{array}\right],
\left[\begin{array}{c}
+1 \\
+1 \\
+1 \\
+1 \\
+1 \\
+1
\end{array}\right],
\end{equation}
and the logical qubit operators
\begin{subequations}
\begin{eqnarray}
X_L & = & XXXXXXX \\
Z_L & = & ZZZZZZZ.
\end{eqnarray}
\end{subequations}
Any single qubit error changes the syndrome.  The syndrome gives
sufficient information to identify which of the qubits received the
error, so it can then be corrected.  This code is a CSS code, and it
has the additional property that it is Hadamard invariant. In
particular, if we apply the Hadamard $H$ operator to every qubit
(``transversal Hadamard''), then the code is unchanged because $H$
performs the transformation
\begin{subequations}
\begin{eqnarray}
H: & X \rightarrow Z \\
   & Z \rightarrow X.
\end{eqnarray}
\end{subequations}
Consequently the stabilizer is transformed into itself with no change
in eigenvalues for the generators.  Furthermore, the transversal
Hadamard exchanges logical $X$ and $Z$ operators, from which we infer
that its effect on the code is a logical Hadamard operator. We call
CSS codes whose set of stabilizers is Hadamard invariant and for which
there are logical operators in CSS form with respect to which the
transversal Hadamard acts as a logical Hadamard on each encoded qubit,
``CSS-$H$ codes''.  Note that $\mathcal{C}_4$ is not a CSS-$H$ code.
The seven-qubit code is the smallest nontrivial CSS-$H$
error-correcting code. However, there is a six-qubit CSS-$H$
error-detecting code encoding two logical qubits, which was described
in \cite{Knill04}. It has generators
\begin{equation}
\mathcal{C}_6 \Leftrightarrow \left[\begin{array}{cccccc}
X & I & X & X & I & X \\
I & X & X & I & X & X \\
Z & I & Z & Z & I & Z \\
I & Z & Z & I & Z & Z
\end{array}\right],
\left[\begin{array}{c}
+1 \\
+1 \\
+1 \\
+1 \\
\end{array}\right],
\end{equation}
and the logical qubit operators
\begin{subequations}
\begin{eqnarray}
X_L^{(1)} & = & XXXIII \\
Z_L^{(1)} & = & ZZZIII \\
X_L^{(2)} & = & IIIXXX \\
Z_L^{(2)} & = & IIIZZZ.
\end{eqnarray}
\end{subequations}
We have chosen different logical operators from those used in
\cite{Knill04}, so that the code is explicitly $H$-invariant.
$\mathcal{C}_6$ detects errors, and if we know whether the error is
in the first or second group of three qubits, we can also correct it.
Otherwise we know what type of error has affected the logical qubits,
but not which logical qubit was affected.  $\mathcal{C}_6$ is the
smallest CSS-$H$ error-detecting code.

\subsection{Operator Arrays}

In entanglement purification we have $n$ parties and $m$ copies of a
large entangled state. The copies are prepared and distributed so that
each party holds one qubit of each copy, for a total of $m$ qubits per
party. The stabilizer group of the full state of this $m\times n$
qubit system has $m \times n$ generators, each of which is composed of
$m \times n$ Pauli matrices.  For example, if
two copies of the Bell pair $|B_{00}\rangle$ are shared by Alice and
Bob, the generators for this four-qubit system are

\begin{equation}
\left[\begin{array}{cccc}
A1 & A2 & B1 & B2 \\
\hline
X & I & X & I \\
Z & I & Z & I \\
I & X & I & X \\
I & Z & I & Z
\end{array}\right],
\label{twobellpairs}
\end{equation}
where we use the top line in the chart to label each qubit as
belonging to Alice or Bob and copy one or two of the shared state.
Notice that qubits $A1$ and $B1$ are entangled with one another but
not with $A2$ and $B2$.

We would like a method for representing the stabilizer generators that
emphasizes the structure of these states.  Instead of writing each
generator as a single row in a table, we write each generator as an
array whose columns belong to a particular party and whose rows
represent a particular copy of the shared entangled state.  The set of
generators is a list of such arrays.  Using these operator
arrays, we represent the two copies of the Bell pair in
Eq. (\ref{twobellpairs}) with
\begin{equation}
\left[\begin{array}{c|cc}
  & A & B \\
\hline
1 & X & X \\
2 & I & I 
\end{array}\right],
\left[\begin{array}{cc}
A & B \\
\hline
Z & Z \\
I & I 
\end{array}\right],
\left[\begin{array}{cc}
A & B \\
\hline
I & I \\
X & X 
\end{array}\right],
\left[\begin{array}{cc}
A & B \\
\hline
I & I \\
Z & Z 
\end{array}\right].
\label{twobellpairsarray}
\end{equation}
Here we can easily see that entanglement exists only within rows; row
one is never entangled with row two. We call the generators of the
state that has been copied (in this case $XX$ and $ZZ$) the ``master
generators''. The members of Eq.~(\ref{twobellpairsarray}) are
``single-copy generators''. We can see that each single-copy generator
has identities in all rows except one, which contains a master
generator. The objects listed in Eq.~(\ref{twobellpairsarray})
generate the ``multi-copy stabilizer''. A useful subset of the multi-copy
stabilizer is the set of ``parallel stabilizer elements,'' which are
products of single-copy generators having the same master
generator. These parallel stabilizer elements have rows equal to the
identity or only one of the master generators.

\section{Error-Correcting Codes for Purification}
\label{errorcorrectingcodesforpurification}
In this section we describe a general method for understanding
entanglement purification protocols using quantum error-correcting
codes. These ideas were described in~\cite{Matsumoto02,Hostens04}. The
essence of the scheme is that each party measures the parity check
operators of an error-correcting code on his or her qubits. By
comparing the results of these measurements the parties learn about
the errors that have corrupted their states. They correct the errors
and then decode logical qubits into the desired entangled state.

To begin a purification protocol, the parties apply random stabilizer
elements to the noisy states they hope to purify. Each party just
applies an agreed-upon single qubit Pauli matrix to his or her qubit of a
particular state. These Pauli matrices make up a random stablizer
element that the parties determine by classical communication before
the start of the protocol.  After doing this, they can treat any noisy
state as a probabilistic mixture of states that have the same
stabilizer but different eigenvalues for the generators. This fact was
proven for Bell states in \cite{Bennett95} and was extended to any
stabilizer state by Aschauer, D\"ur and Briegel in
\cite{Aschauer04}. We include the proof here for pedagogical
completeness.

Let $|\psi_0\rangle$ be the desired state for which all of the
generators eigenvalues are $+1$. Any pure noisy state can be written
as
\begin{equation}
|\psi\rangle=\sum_i \alpha_i P_i |\psi_0\rangle,
\end{equation}
where $i$ extends over all possible syndromes, $P_i$ is a tensor
product of Pauli matrices that moves the state from the all $+1$
syndrome to syndrome $i$, and $\alpha_i$ is an amplitude. If there is
little noise, the $\alpha_i$ are small for $P_i$ not equal to
identity. The density matrix for this state is
\begin{equation}
|\psi\rangle\langle\psi|=\sum_{i,j}\alpha_i \alpha_j^{\ast} P_i |\psi_0\rangle\langle\psi_0|P_j^\dagger.
\end{equation}
An arbitrary noisy mixed state is just a probabilistic mixture of
noisy pure states, which we can write as
\begin{equation}
\rho=\sum_k p_k \sum_{i,j} \alpha_{ik} \alpha_{jk}^{\ast} P_{i}|\psi_0\rangle\langle\psi_0|P_{j}^\dagger,
\end{equation}
where $\sum_k p_k=1$.  Alice and her friends now apply a random
element of the stabilizer group to this state. After ``forgetting'' which
of the $N$ stabilizer elements they applied, the state becomes
\begin{eqnarray}
\rho_Q & = & \frac{1}{N}\sum_{Q\in \mathrm{stab.}}Q\rho Q^\dagger \\
 & = &  \frac{1}{N}\sum_{ijk}p_k\alpha_{ik}\alpha_{jk}^\ast \sum_{Q\in stab}QP_i|\psi_0\rangle\langle \psi_0|P_j^\dagger Q^\dagger.
\end{eqnarray}
Because all of the $Q$'s and $P$'s are made of tensor products of
Pauli matrices, they must either commute or anti-commute with one
another.  Let $\langle Q,P\rangle=0$ if $P$ and $Q$ commute and
$\langle Q,P\rangle=1$ if they anti-commute. After commuting the $P$'s
and $Q$'s we can use the fact that $Q$ is in the stabilizer of
$|\psi_0\rangle$ to obtain
\begin{equation}
\rho_Q=\frac{1}{N} \sum_{ijk}p_k\alpha_{ik}\alpha_{jk}^\ast \sum_{Q\in
\mathrm{stab}} (-1)^{\langle Q,P_i P_j\rangle}
P_i|\psi_0\rangle\langle\psi_0|P_j^\dagger.
\label{eq:randomizedsum}
\end{equation}
Let us examine the sum $\sum_{Q\in\mathrm{stab}} (-1)^{\langle Q,P_i
P_j\rangle}$ over all $Q$ in the stabilizer of $|\psi_0\rangle$ in the
case where $i\not=j$. The elements of the stabilizer that commute
with $P_iP_j$ form a subgroup, $\mathcal{Q}_{ij}$. There must be some
element of the stabilizer that anti-commutes with $P_iP_j$ (otherwise
$P_iP_j$ would itself be in the stabilizer, in which case $i=j$), let
us call this element $q$. We can now generate every element of the
stabilizer that anti-commutes with $P_iP_j$ by multiplying every
element of $\mathcal{Q}_{ij}$ by $q$. (The elements that anti-commute
are a coset of $\mathcal{Q}_{ij}$.) Therefore the number of elements
of the stabilizer that commute with $P_iP_j$ is equal to the number of
elements of the stabilizer that anti-commute with
$P_iP_j$. Consequently the terms of Eq.~(\ref{eq:randomizedsum}) in the
sum over all $Q$ for which $i\ne j$ must all cancel one another
leaving us with
\begin{equation}
\rho_Q=\frac{1}{N} \sum_{ik} p_k |\alpha_{ik}|^2 P_i|\psi_0\rangle\langle\psi_0|P_i^\dagger.
\end{equation}
This is just a mixture, each of whose terms are eigenstates of the
stabilizer operators with different eigenvalues. Therefore it would be
sufficient to measure each of the generators to extract an error
syndrome, correct errors and, if all the errors are corrected, obtain a
pure state. However, because each party holds only a single qubit,
they cannot measure the generators directly.

In the picture of the operator arrays, each party can measure any
operator that has non-identity entries only along his or her column,
but the master generators describing the state are oriented along
rows. We can use techniques of error-correcting codes to overcome this
problem. Each party will use an error-detecting or -correcting code that
encodes one or more logical qubits on the $m$ physical
qubits. We will assume that every party will use the
same code. They each measure the generators of that code and then
share the measurement results. Depending on the construction of the
code and the original state they are trying to purify, they expect a
particular pattern of correlations between their measurement
results. Errors in the transmission of the $m$ entangled states should
appear as aberrations in the syndrome patterns, so they can be
detected or corrected.  The parties then have an encoded copy of a
state, which they can decode. 

Let us consider an example. Suppose Alice and Bob wish to purify Bell
pairs $|B_{00}\rangle$, and they share four copies of noisy Bell
pairs.  We first examine the case in which no errors are present.  The
single-copy generators of the full eight qubit state are
\begin{eqnarray}
\left[\begin{array}{c|cc}
  & A & B \\
\hline
1 & X & X \\
2 & I & I \\
3 & I & I \\
4 & I & I \\
\end{array}\right],
\left[\begin{array}{cc}
A & B \\
\hline
Z & Z \\
I & I \\
I & I \\
I & I \\
\end{array}\right],
\left[\begin{array}{cc}
A & B \\
\hline
I & I \\
X & X \\
I & I \\
I & I \\
\end{array}\right],
\left[\begin{array}{cc}
A & B \\
\hline
I & I \\
Z & Z \\
I & I \\
I & I \\
\end{array}\right], \nonumber \\
\left[\begin{array}{c|cc}
  & A & B \\
\hline
1 & I & I \\
2 & I & I \\
3 & X & X \\
4 & I & I \\
\end{array}\right],
\left[\begin{array}{cc}
A & B \\
\hline
I & I \\
I & I \\
Z & Z \\
I & I \\
\end{array}\right],
\left[\begin{array}{cc}
A & B \\
\hline
I & I \\
I & I \\
I & I \\
X & X \\
\end{array}\right],
\left[\begin{array}{cc}
A & B \\
\hline
I & I \\
I & I \\
I & I \\
Z & Z \\
\end{array}\right].
\label{fourbellpairs}\end{eqnarray}
Alice and Bob can use the $\mathcal{C}_4$ code to purify, so they each
measure the two generators of $\mathcal{C}_4$ on his and her own
qubits.  The stabilizer arrays describing these single-party parity
checks are
\begin{eqnarray}
\left[\begin{array}{c|cc}
  & A & B \\
\hline
1 & X & I \\
2 & X & I \\
3 & X & I \\
4 & X & I \\
\end{array}\right],
\left[\begin{array}{cc}
A & B \\
\hline
Z & I \\
Z & I \\
Z & I \\
Z & I \\
\end{array}\right],
\left[\begin{array}{cc}
A & B \\
\hline
I & X \\
I & X \\
I & X \\
I & X \\
\end{array}\right],
\left[\begin{array}{cc}
A & B \\
\hline
I & Z \\
I & Z \\
I & Z \\
I & Z \\
\end{array}\right].
\label{c4measurements}
\end{eqnarray}
The result of each single-party parity check will be $\pm 1$ with
probability $\frac{1}{2}$. However
\begin{eqnarray}
\left[\begin{array}{c|cc}
  & A & B \\
\hline
1 & X & X \\
2 & X & X \\
3 & X & X \\
4 & X & X \\
\end{array}\right]
\mathrm{and}
\left[\begin{array}{cc}
A & B \\
\hline
Z & Z \\
Z & Z \\
Z & Z \\
Z & Z \\
\end{array}\right]
\label{eightxszs}
\end{eqnarray}
are in the multi-copy stabilizer of the states Alice and Bob are
purifying. The operators given in Eq.~(\ref{eightxszs}) are parallel
stabilizer elements made by repeating the master generators on
multiple rows. They are also examples of ``parallel parity checks'' in
which the same generator of the purifying code is repeated on multiple
columns. In the absence of error, the eigenvalues of these operators
are $+1$ (because they are in the stabilizer), so the eigenvalues
Alice and Bob obtain for their $XXXX$ measurements must match and
similarly for their $ZZZZ$ measurements. Let us assume that after
these measurements, Alice and Bob transform their states by applying
known Pauli products so that they all have $+1$ eigenvalues for the
operators in Eq.~(\ref{c4measurements}).  Note that this is not
strictly necessary as long as Alice and Bob keep track of their Pauli
frame, where the Pauli frame is defined by a Pauli product that would
restore the system so that all stabilizer generators have eigenvalue
$+1$. If Alice and Bob know the Pauli frame they can simply adjust
future manipulations of their states to compensate for the changes
in eigenvalues without ever applying Pauli product compensations.

We can now find the generators of the new stabilizer that Alice and
Bob share after their measurements. The new stabilizer must include
all of the measurement operators and all elements of the old
stabilizer that commute with the measurements. Eight generators
are required and they include all of Eq.~(\ref{c4measurements}) and
\begin{eqnarray}
\left[\begin{array}{c|cc}
  & A & B \\
\hline
1 & X & X \\
2 & X & X \\
3 & I & I \\
4 & I & I \\
\end{array}\right],
\left[\begin{array}{cc}
A & B \\
\hline
Z & Z \\
I & I \\
Z & Z \\
I & I \\
\end{array}\right],
\left[\begin{array}{cc}
A & B \\
\hline
I & I \\
X & X \\
I & I \\
X & X \\
\end{array}\right],
\left[\begin{array}{cc}
A & B \\
\hline
I & I \\
I & I \\
Z & Z \\
Z & Z \\
\end{array}\right].
\label{bellpairsinc4}
\end{eqnarray}
These are parallel stabilizer elements, so they are in the
original multi-copy stabilizer. Alice and Bob now each possess two
logical qubits encoded in $\mathcal{C}_4$. We can see the state of
these logical qubits using the logical encoded operators given in
Eq.~(\ref{c4logical}). We rewrite the operators in
Eq.~(\ref{bellpairsinc4}) using the logical qubit operators as
\begin{eqnarray}
\left[\begin{array}{c|cc}
  & A & B \\
\hline
1 & X_L & X_L \\
2 & I & I \\
\end{array}\right],
\left[\begin{array}{cc}
A & B \\
\hline
Z_L & Z_L \\
I & I \\
\end{array}\right],
\left[\begin{array}{cc}
A & B \\
\hline
I & I \\
X_L & X_L \\
\end{array}\right],
\left[\begin{array}{cc}
A & B \\
\hline
I & I \\
Z_L & Z_L \\
\end{array}\right].
\end{eqnarray}
These are the master generators for two Bell pairs shared between
Alice and Bob -- exactly the state they wanted to purify.

Let us now examine the behavior of this scheme in the presence of an
error. An error changes the eigenvalue of one of the
single copy generators in Eq.~(\ref{fourbellpairs}) to $-1$. Alice
and Bob detect this change when they use their single-party
parity checks to compile the multi-party parity checks. For example,
if Alice's first qubit suffers from a $Z$ error, the eigenvalue of the
first single-copy generator in Eq.~(\ref{fourbellpairs})'s list is
$-1$. Also, the eigenvalue of the first operator in
Eq.~(\ref{eightxszs})'s list is $-1$. Therefore the product of the
eigenvalues Alice and Bob obtain from their $XXXX$ single-party parity
checks must be $-1$. They detect this error when they compare
measurement results and obtain their multi-party parity check. A $Z$
error to any of the eight qubits will cause this same error syndrome,
so Alice and Bob cannot correct it.

This analysis allows us to formulate sufficient
conditions on the success of purification schemes of this form: (1)
The multi-party parity checks that the parties perform must be
sensitive to any change in the eigenvalues of the generators of the
states they wish to purify. (2) The stabilizer of the desired encoded
state must be in the original multi-copy stabilizer of the qubits to
be purified.

\section{Purifying Stabilizer States with Stabilizer Codes}
\label{statesandcodes}

In this section we discuss which classes of states may be purified
using specified classes of error-correcting codes. The simplest class of
states are the CSS-$H$ states, which can be transformed using local
operations into states whose stabilizer generators can be written in
CSS form and are $H$ invariant as a set. More complex states are CSS,
but not $H$ invariant. The most general class of states we consider
includes all stabilizer states. We similarly classify codes as being
CSS-$H$, CSS, or any stabilizer code.

\subsection{Purifying CSS-$H$ States}

All CSS-$H$ states must contain an even number of qubits because
there are equal numbers of $Z$-type and $X$-type generators.  The only
two, four and six qubit CSS-$H$ states are collections of Bell pairs,
but more complicated states can be formed from eight or more
qubits. In the following we describe how any CSS-$H$ state can be
purified by use of any error-detecting stabilizer code. We do this by
showing that the multi-copy stabilizer contains enough versatility to
allow any error-detecting stabilizer code to meet the conditions (1)
and (2) stated an the end of the previous section. Matsumoto has
already demonstrated that any stabilizer code can be used to purify
Bell pairs and maximally entangled bipartite states of qudits
\cite{Matsumoto02}.

Let us examine the example of Alice and Bob purifying a Bell pair
using $\mathcal{C}_5$ as their purifying code. The Bell pair has
master generators $XX$ and $ZZ$, and the multi-copy stabilizer is
generated by the single-copy generators
\begin{eqnarray}
\left[\begin{array}{c|cc}
  & A & B \\
\hline
1 & X & X \\
2 & I & I \\
3 & I & I \\
4 & I & I \\
5 & I & I \\
\end{array}\right],
\left[\begin{array}{cc}
A & B \\
\hline
Z & Z \\
I & I \\
I & I \\
I & I \\
I & I \\
\end{array}\right],
\left[\begin{array}{cc}
A & B \\
\hline
I & I \\
X & X \\
I & I \\
I & I \\
I & I \\
\end{array}\right],
\left[\begin{array}{cc}
A & B \\
\hline
I & I \\
Z & Z \\
I & I \\
I & I \\
I & I
\end{array}\right],
\left[\begin{array}{cc}
A & B \\
\hline
I & I \\
I & I \\
X & X \\
I & I \\
I & I 
\end{array}\right], \nonumber \\
\left[\begin{array}{c|cc}
  & A & B \\
\hline
1 & I & I \\
2 & I & I \\
3 & Z & Z \\
4 & I & I \\
5 & I & I 
\end{array}\right],
\left[\begin{array}{cc}
A & B \\
\hline
I & I \\
I & I \\
I & I \\
X & X \\
I & I 
\end{array}\right],
\left[\begin{array}{cc}
A & B \\
\hline
I & I \\
I & I \\
I & I \\
Z & Z \\
I & I
\end{array}\right],
\left[\begin{array}{cc}
A & B \\
\hline
I & I \\
I & I \\
I & I \\
I & I \\
X & X
\end{array}\right],
\left[\begin{array}{cc}
A & B \\
\hline
I & I \\
I & I \\
I & I \\
I & I \\
Z & Z
\end{array}\right].
\label{fivebellpairs}\end{eqnarray}
Alice and Bob obtain single-party parity checks by measuring
\begin{eqnarray}
\left[\begin{array}{c|cc}
  & A & B \\
\hline
1 & X & I \\
2 & Z & I \\
3 & Z & I \\
4 & X & I \\
5 & I & I \\
\end{array}\right],
\left[\begin{array}{cc}
A & B \\
\hline
I & I \\
X & I \\
Z & I \\
Z & I \\
X & I \\
\end{array}\right],
\left[\begin{array}{cc}
A & B \\
\hline
X & I \\
I & I \\
X & I \\
Z & I \\
Z & I \\
\end{array}\right],
\left[\begin{array}{cc}
A & B \\
\hline
Z & I \\
X & I \\
I & I \\
X & I \\
Z & I
\end{array}\right], \nonumber \\
\left[\begin{array}{c|cc}
  & A & B \\
\hline
1 & I & X \\
2 & I & Z \\
3 & I & Z \\
4 & I & X \\
5 & I & I 
\end{array}\right],
\left[\begin{array}{cc}
A & B \\
\hline
I & I \\
I & X \\
I & Z \\
I & Z \\
I & X 
\end{array}\right],
\left[\begin{array}{cc}
A & B \\
\hline
I & X \\
I & I \\
I & X \\
I & Z \\
I & Z 
\end{array}\right],
\left[\begin{array}{cc}
A & B \\
\hline
I & Z \\
I & X \\
I & I \\
I & X \\
I & Z
\end{array}\right].
\end{eqnarray}
Alice and Bob must now construct multi-party parity checks that are in
the multi-party stabilizer and are sensitive to any error that would
change the eigenvalue of any of the single-copy generators. They can
do this by multiplying their single-party parity checks in parallel,
so that they obtain the eigenvalues of
\begin{equation}
\left[\begin{array}{c|cc}
  & A & B \\
\hline
1 & X & X \\
2 & Z & Z \\
3 & Z & Z \\
4 & X & X \\
5 & I & I 
\end{array}\right],
\left[\begin{array}{cc}
A & B \\
\hline
I & I \\
X & X \\
Z & Z \\
Z & Z \\
X & X 
\end{array}\right],
\left[\begin{array}{cc}
A & B \\
\hline
X & X \\
I & I \\
X & X \\
Z & Z \\
Z & Z 
\end{array}\right],
\left[\begin{array}{cc}
A & B \\
\hline
Z & Z \\
X & X \\
I & I \\
X & X \\
Z & Z 
\end{array}\right].
\label{c5measurements}
\end{equation}
Each of these operators is in the multi-copy stabilizer because each
has rows equal to the master generators.

If no errors have occurred, Alice and Bob find that all of the
multi-party parity checks give eigenvalues of $+1$. Suppose for
example that Alice's first qubit has suffered from a $Z$ error. Then
the eigenvalue of the first member of Eq.~(\ref{fivebellpairs}) will
be $-1$. When Alice and Bob examine their multi-party parity checks,
they find that the first and third members of
Eq.~(\ref{c5measurements}) have eigenvalue $-1$. In practice it is not
even necessary to identify the particular qubit that suffered the
error.  They need only to know how to correctly return the encoded
space to the subsystem with all $+1$ eigenvalues for the generators of
the stabilizers.  They can therefore choose to apply $Z$ to Alice's or Bob's
first qubit to correct this error. Any single qubit error will give a
different syndrome pattern. However, multi-qubit errors (such as an
$X$ error to Alice's second qubit and an $X$ error to Bob's fifth
qubit) can result in the same syndromes as single qubit errors and
would cause Alice and Bob to incorrectly ``correct'' the
errors. However, any combination of errors restricted to a single row
of the array can be corrected.  Depending on their error-model and the
particular syndrome result they obtain, Alice and Bob may decide to
simply discard their states and try again.

The $X$ and $Z$ Pauli operators for Alice's and Bob's encoded qubits are
\begin{eqnarray}
X_L^{(A)}=\left[\begin{array}{c|cc}
  & A & B \\
\hline
1 & X & I \\
2 & X & I \\
3 & X & I \\
4 & X & I \\
5 & X & I 
\end{array}\right], &
Z_L^{(A)}=\left[\begin{array}{cc}
A & B \\
\hline
Z & I \\
Z & I \\
Z & I \\
Z & I \\
Z & I 
\end{array}\right], \nonumber\\
X_L^{(B)}=\left[\begin{array}{c|cc}
  & A & B \\
\hline
1 & I & X \\
2 & I & X \\
3 & I & X \\
4 & I & X \\
5 & I & X
\end{array}\right], &
Z_L^{(B)}=\left[\begin{array}{cc}
A & B \\
\hline
I & Z \\
I & Z \\
I & Z \\
I & Z \\
I & Z
\end{array}\right].
\end{eqnarray}
Alice and Bob would like to have purified the state whose encoded
generators are the master generators, i.e.
\begin{equation}
X_L^{(A)}X_L^{(B)}=\left[\begin{array}{c|cc}
  & A & B \\
\hline
1 & X & X \\
2 & X & X \\
3 & X & X \\
4 & X & X \\
5 & X & X
\end{array}\right], \quad\mathrm{and}\quad
Z_L^{(A)}Z_L^{(B)}=\left[\begin{array}{cc}
A & B \\
\hline
Z & Z \\
Z & Z \\
Z & Z \\
Z & Z \\
Z & Z
\end{array}\right].
\end{equation} 
They will have this state after projecting the multi-copy state into
the encoded subspace by measuring the code's generators, provided that
the encoded master generators were in the original multi-copy
stabilizer. This is surely the case because the rows of the encoded
master generators' arrays contain only the master generators.
 
Any other CSS-$H$ state can be purified in a similar manner. We need
only specify a method for constructing the multi-party parity checks
from the single-party parity checks. Each multi-party parity check
should include columns that contain the identity or only one of the
generators of the purifying code. Which columns have identity and
which contain the generator of the purifying code are determined so
that the rows of the multi-party parity check array match a particular
master generator (containing $X$'s), its Hadamard-pair (containing
$Z$'s), or the product of a master generator and its Hadamard pair
(containing $Y$'s). For each Hadamard-pair of master generators we
construct a number of multi-party parity checks equal to the number of
generators for the purifying code (four for $\mathcal{C}_5$). These
generators give a syndrome that tells us which copy has received an
$X$, $Y$, or $Z$ error affecting that Hadamard-pair of master
generators. Similar syndromes are obtained for each pair of master
generators. We now know which copy of the original state received an
error and how that error affected each of that copy's pairs of
generators. We can use this information to determine a Pauli product
to apply to this copy to restore the correct syndrome and thus fix the
error.  For each Hadamard pair of master generators, we can correct an
error that affects only one Hadamard pair of single-copy
generators. Multiple errors may be corrected provided that they each
affect single-copy generators associated with different Hadamard pairs
of master generators. If qubits in multiple rows receive errors
affecting the same Hadamard pair of master generators, this procedure
may be confused, so Alice and her friends may want to use a more
powerful error-correcting code.

The encoded master generator must also be in the multi-copy
stabilizer. We can see that this is always true because each master
generator contains only $X$'s or $Z$'s (and $I$'s), and each master
generator has a Hadamard pair. An encoded generator's array
contains columns corresponding to the encoded $X_L$ (or $Z_L$)
operators or the identity, and each row therefore contains that
particular master generator, its Hadamard pair, or the identity. Every
array whose rows are master generators or the identity are in the
multi-copy stabilizer, thus the encoded master generator is in the
multi-copy stabilizer. Therefore any stabilizer code can be used to
purify a CSS-$H$ state.

\subsection{Purifying CSS States}

Methods for purifying any multi-party CSS state have been obtained by
Aschauer, D\"ur and Briegel in \cite{Aschauer04} and by Hostens,
Dehaene and De Moor in \cite{Hostens05}. Our goal here is to show
that any CSS state can be purified by use of any CSS error-detecting or
-correcting code.

Let us use for an example the three qubit GHZ state with master
generators $ZZI$, $XXX$ and $IZZ$. This is the simplest CSS state
that is not $H$ invariant and has more than one qubit. Its graph is a
three node path. It is instructive to consider why the non-CSS code
$\mathcal{C}_5$ is unable to purify this state. Suppose that five
copies of this state are distributed to Alice, Bob and Charlie. The
multi-copy stabilizer is generated by
\begin{eqnarray}
\left[\begin{array}{c|ccc}
  & A & B & C \\
\hline
1 & Z & Z & I \\
2 & I & I & I \\
3 & I & I & I \\
4 & I & I & I \\
5 & I & I & I
\end{array}\right], &
\left[\begin{array}{ccc}
A & B & C \\
\hline
X & X & X \\
I & I & I \\
I & I & I \\
I & I & I \\
I & I & I
\end{array}\right], &
\left[\begin{array}{ccc}
A & B & C \\
\hline
I & Z & Z \\
I & I & I \\
I & I & I \\
I & I & I \\
I & I & I
\end{array}\right], \nonumber\\
\vdots & \vdots & \vdots \nonumber \\
\left[\begin{array}{c|ccc}
  & A & B & C \\
\hline
1 & I & I & I \\
2 & I & I & I \\
3 & I & I & I \\
4 & I & I & I \\
5 & Z & Z & I
\end{array}\right], & 
\left[\begin{array}{ccc}
A & B & C \\
\hline
I & I & I \\
I & I & I \\
I & I & I \\
I & I & I \\
X & X & X
\end{array}\right], &
\left[\begin{array}{ccc}
A & B & C \\
\hline
I & I & I \\
I & I & I \\
I & I & I \\
I & I & I \\
I & Z & Z
\end{array}\right].
\end{eqnarray}
Using $\mathcal{C}_5$ as their purifying code, Alice, Bob and Charlie
measure the single-party parity checks
\begin{eqnarray}
\left[\begin{array}{c|ccc}
  & A & B & C \\
\hline
1 & X & I & I \\
2 & Z & I & I \\
3 & Z & I & I \\
4 & X & I & I \\
5 & I & I & I
\end{array}\right], &
\left[\begin{array}{ccc}
A & B & C \\
\hline
I & I & I \\
X & I & I \\
Z & I & I \\
Z & I & I \\
X & I & I
\end{array}\right], &
\left[\begin{array}{ccc}
A & B & C \\
\hline
X & I & I \\
I & I & I \\
X & I & I \\
Z & I & I \\
Z & I & I
\end{array}\right], \nonumber\\
\left[\begin{array}{c|ccc}
  & A & B & C \\
\hline
1 & Z & I & I \\
2 & X & I & I \\
3 & I & I & I \\
4 & X & I & I \\
5 & Z & I & I
\end{array}\right],&
\hdots,&
\left[\begin{array}{ccc}
A & B & C \\
\hline
I & I & Z \\
I & I & X \\
I & I & I \\
I & I & X \\
I & I & Z
\end{array}\right].
\end{eqnarray}
Using these arrays Alice, Bob and Charlie must now construct
multi-party parity checks that are in the multi-copy stabilizer and
sensitive to the eigenvalues of all of the generators of the
multi-copy stabilizer. How can they check the eigenvalue of the
operator $ZZI$ acting on the first copy? They might attempt to combine
Alice's and Bob's measurement of $ZXIXZ$ to produce the multi-party
parity check
\begin{equation}
\left[\begin{array}{c|ccc}
  & A & B & C \\
\hline
1 & Z & Z & I \\
2 & X & X & I \\
3 & I & I & I \\
4 & X & X & I \\
5 & Z & Z & I
\end{array}\right].
\end{equation}
This is unfortunately not in the multi-copy stabilizer because
the second row, $XXI$, is neither one of the master generators nor a
product of some of them. The result of this measurement will then be $\pm 1$
with probability $\frac{1}{2}$, regardless of any errors. In fact it is
not possible for Alice, Bob and Charlie to construct a set of
multi-party parity checks using this code, which is sufficient for detecting
errors, so their attempt at purification fails. However, in the
absence of errors, in this particular case the encoded state is still
a GHZ state.

Suppose instead that Alice, Bob and Charlie use a CSS code such as
$\mathcal{C}_4$. Then their single-party parity checks are columns
containing only $X$'s or $Z$'s. They can combine parallel single-party
parity checks (by which they have measured the same generator of the
purifying code) to form multi-party parity checks whose rows are
repetitions of the same master generator or the identity. These are
parallel elements of the multi-copy stabilizer, and by construction of
the error-detecting (or -correcting) code they can detect (or correct)
some set of errors on the initial states. For each master generator
containing $X$'s we have multi-party parity checks for each generator
of the purifying code that contains $X$'s. This gives a particular
syndrome pattern for diagnosing errors of that master generator on
each copy. If the purifying code is one-error-correcting, we can tell
which copy has received a $Z$ (or $Y$) error affecting that master
generator. Each master generator containing $Z$'s is also matched with
a particular syndrome pattern by measuring the code's $Z$-containing
generators. This can tell us which copy has received an $X$ (or $Y$)
affecting that master generator. When using a one-error-correcting
code, this scheme can correct a single error affecting one
single-copy generator associated with each of the master
generators. Multiple errors can be corrected provided that they each
affect single-copy generators associated with different master
generators. If the error is correctable, the parties can determine a
Pauli product to correct it.

The encoded logical operators also contain only $X$'s or $Z$'s and
they can similarly be combined in parallel to match the master
generators, so they are also in the multi-copy stabilizer. This
provides the method to purify any CSS state with any CSS code.

\subsection{Purifying Any Stabilizer State}

One can purify any stabilizer state in a manner similar to that just
described for CSS-$H$ states and CSS states; we just need to find an
appropriate class of error-detecting or -correcting codes.  For
an independent approach to purifying any stabilizer state, see
\cite{Kruszynska06}.

Let us use the three qubit state with master generators $XZZ$, $ZXZ$,
and $ZZX$ as an example. The graph of this state is a triangle, and it
is impossible to transform it into a CSS state by use of single qubit
operations. The multi-copy stabilizer for this state includes all
arrays with rows given by the master stabilizers. Alice, Bob and
Charlie must ensure that their purifying code can produce multi-party
parity checks and encoded master generators that are in the multi-copy
stabilizer. They cannot use $\mathcal{C}_5$ because there is no method
of combining the single-party parity checks to produce a non-trivial
element of the multi-copy stabilizer. They also cannot use
$\mathcal{C}_4$ because the encoded logical Pauli operators form
arrays whose rows are not equal to master generators and are therefore
not in the multi-copy stabilizer.

Suppose Alice, Bob and Charlie use $\mathcal{C}_7$ as their purifying
code. The multi-copy stabilizer has the generators
\begin{eqnarray}
\left[\begin{array}{c|ccc}
  & A & B & C \\
\hline
1 & X & Z & Z \\
2 & I & I & I \\
\vdots & \vdots & \vdots & \vdots \\
7 & I & I & I
\end{array}\right], &
\left[\begin{array}{ccc}
A & B & C \\
\hline
Z & X & Z \\
I & I & I \\
\vdots & \vdots & \vdots \\
I & I & I
\end{array}\right], &
\left[\begin{array}{ccc}
A & B & C \\
\hline
Z & Z & X \\
I & I & I \\
\vdots & \vdots & \vdots \\
I & I & I
\end{array}\right], \nonumber\\
\vdots & \vdots & \vdots \nonumber \\
\left[\begin{array}{c|ccc}
  & A & B & C \\
\hline
1 & I & I & I \\
\vdots & \vdots & \vdots & \vdots \\
6 & I & I & I \\
7 & X & Z & Z
\end{array}\right], & 
\left[\begin{array}{ccc}
A & B & C \\
\hline
I & I & I \\
\vdots & \vdots & \vdots \\
I & I & I \\
Z & X & Z
\end{array}\right], &
\left[\begin{array}{ccc}
A & B & C \\
\hline
I & I & I \\
\vdots & \vdots & \vdots \\
I & I & I \\
Z & Z & X
\end{array}\right].
\end{eqnarray}
Alice measures the single-party parity checks
\begin{eqnarray}
\left[\begin{array}{c|ccc}
  & A & B & C \\
\hline
1 & X & I & I \\
2 & I & I & I \\
3 & X & I & I \\
4 & I & I & I \\
5 & X & I & I \\
6 & I & I & I \\
7 & X & I & I \\
\end{array}\right],
\left[\begin{array}{ccc}
A & B & C \\
\hline
I & I & I \\
X & I & I \\
X & I & I \\
I & I & I \\
I & I & I \\
X & I & I \\
X & I & I \\
\end{array}\right],
\left[\begin{array}{ccc}
A & B & C \\
\hline
I & I & I \\
I & I & I \\
I & I & I \\
X & I & I \\
X & I & I \\
X & I & I \\
X & I & I \\
\end{array}\right], \nonumber \\
\left[\begin{array}{c|ccc}
  & A & B & C \\
\hline
1 & Z & I & I \\
2 & I & I & I \\
3 & Z & I & I \\
4 & I & I & I \\
5 & Z & I & I \\
6 & I & I & I \\
7 & Z & I & I \\
\end{array}\right],
\left[\begin{array}{ccc}
A & B & C \\
\hline
I & I & I \\
Z & I & I \\
Z & I & I \\
I & I & I \\
I & I & I \\
Z & I & I \\
Z & I & I \\
\end{array}\right],
\left[\begin{array}{ccc}
A & B & C \\
\hline
I & I & I \\
I & I & I \\
I & I & I \\
Z & I & I \\
Z & I & I \\
Z & I & I \\
Z & I & I \\
\end{array}\right].
\end{eqnarray}
Bob and Charlie obtain similar parity checks, except that the
non-identity operators are shifted to Bob's and Charlie's
columns. Each single-party parity check has a Hadamard pair. Alice,
Bob and Charlie can obtain multi-party parity checks by combining in
parallel the same measurements or their $Y$ or $Z$ variants to match a
single master generator repeated on several rows.  For example, they
can check the parity of
\begin{equation}
\left[\begin{array}{c|ccc}
  & A & B & C \\
\hline
1 & X & Z & Z \\
2 & I & I & I \\
3 & X & Z & Z \\
4 & I & I & I \\
5 & X & Z & Z \\
6 & I & I & I \\
7 & X & Z & Z \\
\end{array}\right]
\end{equation}
using Alice's measurement of $XIXIXIX$, Bob's measurement of
$ZIZIZIZ$ and Charlie's measurement of $ZIZIZIZ$. They can use this
method to obtain three multi-party parity checks for each master
generator because $\mathcal{C}_7$ has three Hadamard pairs of
generators. If they were trying to purify a state with $Y$'s in the
master generators, they could make multi-party parity checks with
columns of $Y$'s by use of products of Hadamard pairs. This is sufficient
for Alice, Bob and Charlie to locate and correct any single qubit
error.

Now Alice, Bob and Charlie each have a single qubit encoded in the
$\mathcal{C}_7$ subspace, and this qubit has the encoded logical
operators $X_L^{(A,B,\mathrm{\ or\ }C)}=XXXXXXX$ and
$Z_L^{(A,B,\mathrm{\ or\ }C)}=ZZZZZZZ$ oriented in columns. $X_L$ and
$Z_L$ are a Hadamard pair. The encoded master generators have arrays
that contain $X_L$'s and $Z_L$'s in columns, and they match a master
generator in each row. Therefore the encoded master generators are in
the multi-copy stabilizer, and Alice, Bob and Charlie can use
$\mathcal{C}_7$ to purify this state.

Alice and her friends with alphabetized names can use any
error-detecting or -correcting CSS-$H$ code to purify any stabilizer
state of many qubits. The smallest example is the error-detecting code
$\mathcal{C}_6$, which purifies two copies of the desired state from
six.  The multi-party parity checks are built from Hadamard variants
of the code's generators to match each of the master generators of the
state they wish to purify. For each master generator they construct a
number of multi-party parity checks equal to half of the number of
generators of the purifying code. If the code can correct errors,
these parity checks give a syndrome that is sufficient to identify
which copy suffered an error affecting that master generator. They
repeat the procedure diagnosing a syndrome for each master generator,
so that any single qubit error is at least detected. Any error
combination that affects only one single-copy generator in each set of
single-copy generators associated with a single master generator can
be corrected by a one-error-correcting code.

If their code is an effective error-detecting or -correcting code,
they can detect or correct (insofar as the code is able to correct)
errors on their qubits. The encoded master generators are formed from
Hadamard variants of encoded logic operators, so they are present in
the multi-copy stabilizer.

\section{Conclusions and Discussion}
\label{conclusion}

We introduced a method for understanding entanglement
purification with stabilizer codes using operator arrays. We explained
how one can purify (1) any CSS Hadamard invariant state using any
error-detecting stabilizer code, (2) any CSS state using any
error-detecting CSS code and (3) any stabilizer state using any
error-detecting CSS Hadamard invariant code. The smallest code that
can purify any stabilizer state is $\mathcal{C}_6$, which is an
error-detecting code encoding two logical qubits in six.  The smallest
error-correcting code that can purify any stabilizer state is 
Steane's code $\mathcal{C}_7$, which encodes one logical qubit in
seven. These state/code combinations are sufficient for purification,
because they ensure that the parties purifying the state can construct
multi-party parity checks that are in the stabilizer of the copies of the state they hope to purify, and the generators describing
the desired encoded state are also in the original stabilizer.  We
expect that the results we described here can be extended to purify
states of $d$-dimensional quantum systems using stabilizer codes
such as those described in \cite{Gottesman98} and the appropriate
generalizations of CSS and CSS-H codes.

These results raise many questions. For example, it is clear that
there is great freedom in choosing codes to purify states. While we
have given some sufficient conditions for choosing codes, we have not
studied how to match codes and states to give high efficiency of
purified state production. We expect each state/code combination to
have its own conditions on the required fidelity of the input
states. Strategies for maximizing thresholds and efficiencies are
likely to be as rich as those used in other fault-tolerant quantum
information processing tasks.

We anticipate that the entanglement purification methods we have
described using stabilizer codes can be translated into the languages of
permutation or hashing protocols and the graph state methods of
\cite{Kruszynska06}.  This might be accomplished using the lexicon
given by Hostens, Dehaene and De Moor in \cite{Hostens04}. Such a
translation may deepen our understanding of all of these methods.

The scheme we outlined here may also have uses that extend beyond
entanglement purification. Some interesting effects can occur even
when we choose state/code combinations that fail. For example if we
try to purify two copies of the non-CSS triangle state with stabilizer
generators $XZZ$, $ZXZ$ and $ZZX$ using the code $\mathcal{C}_4$, we
can detect any errors to the multi-copy stabilizer because the parity
checks are of CSS-$H$ form. However, the encoded logical Pauli
operators are not. The state obtained after the ``purification'' is
not two copies of the triangle state, but instead an encoded entangled
six qubit CSS state whose graph is shaped like a hexagon. With clever
choices for states and codes we can use this procedure for a type of
remote state preparation that includes error-detection and -correction
capabilities.

\begin{acknowledgments}
We thank John Jost and Ryan Epstein for helpful comments on this
manuscript. This paper is a contribution by the National Institute of
Standards and Technology and not subject to US copyright.
\end{acknowledgments}

\end{document}